\documentclass[useAMS,usenatbib]{mn2e}
\usepackage{graphics,epsf,graphicx,epsf,longtable,amssymb}
\usepackage[T1]{fontenc}
\usepackage{mathptmx}

\def \NII {\ifmmode{[{\rm N{\sc ii}}]~}\else{[N{\sc ii}]~}\fi}
\def \FEII {\ifmmode{[{\rm Fe{\sc ii}}]~}\else{[Fe{\sc ii}]~}\fi}
\def \HA {\ifmmode{{\rm H}\alpha~}\else{${\rm H}\alpha$~}\fi}
\def \kms{\ifmmode{{\rm km\,s}^{-1}}\else{km~s$^{-1}$}\fi}

\title[High-speed knots in Hb~12]{High-speed knots in the hourglass shaped planetary nebula Hubble~12}

\author[Vaytet et al.]{N.~M.~H. Vaytet$^{1}$\footnotemark[1], A.~P. Rushton$^{2}$, M. Lloyd$^{2}$, J.~A. L\'{o}pez$^{3}$, J. Meaburn$^{2}$, \newauthor T.~J. O'Brien$^{2}$, D.~L. Mitchell$^{2}$ and D. Pollacco$^{4}$\\~\\
$^{1}$Service d'Astrophysique, CEA/DSM/DAPNIA/SAp, Centre d'\'{E}tudes de Saclay, L'Orme des Merisiers, 91191 Gif-sur-Yvette,\\ \hspace{20pt} Cedex, France\\
$^{2}$Jodrell Bank Centre for Astrophysics, Alan Turing Building, School of Physics and Astronomy, The University of Manchester,\\ \hspace{20pt} Manchester, M13 9PL, UK\\
$^{3}$Instituto de Astronom\'{i}a, Universidad Nacional Autnoma de M\'{e}xico, Apartado Postal 877, 22800 Ensenada, B.C., M\'{e}xico\\
$^{4}$Department of Pure and Applied Physics, Queen's University Belfast, Belfast, BT7 1NN, UK}

\begin{document}

\date{Received... Accepted...}

\pagerange{\pageref{firstpage}--\pageref{lastpage}} \pubyear{2009}

\maketitle

\label{firstpage}

\begin{abstract}
We present a detailed kinematical analysis of the young compact hourglass-shaped planetary nebula Hb~12. We performed optical imaging and longslit spectroscopy of Hb~12 using the Manchester echelle spectrometer with the 2.1m San Pedro M\'{a}rtir telescope. We reveal, for the first time, the presence of end caps (or knots) aligned with the bipolar lobes of the planetary nebula shell in a deep \NII$\lambda6584$ image of Hb~12. We measured from our spectroscopy radial velocities of $\sim120~\kms$ for these knots.

We have derived the inclination angle of the hourglass shaped nebular shell to be $\sim 65^{\circ}$ to the line of sight. It has been suggested that Hb~12's central star system is an eclipsing binary \citep{hsia06} which would imply a binary inclination of at least $80^{\circ}$. However, if the central binary has been the major shaping influence on the nebula then both nebula and binary would be expected to share a common inclination angle.

Finally, we report the discovery of high-velocity knots with Hubble-type velocities, close to the core of Hb~12, observed in \HA and oriented in the same direction as the end caps. Very different velocities and kinematical ages were calculated for the outer and inner knots showing that they may originate from different outburst events.
\end{abstract}


\footnotetext[1]{{\it E-mail}: neil.vaytet@cea.fr}

\section{Introduction}\label{s1}

Understanding the wide variety of planetary nebulae (PNe) morphologies has puzzled the stellar evolution community for many years. The Generalized Interacting Stellar Winds (GISW) model has long been the widely accepted model for the production of non-spherical PNe (\citealt*{kahn85,balick87}). It involves a strong equatorial density enhancement in the red giant wind around the progenitor, into which runs the fast wind from the central white dwarf, which causes the fast wind to progress faster in the polar direction, thus giving the nebula its bipolar shape.

However, in recent years, \citet{sahai98} and \citet{soker00} have argued that extremely elongated nebulae and bipolar shells with very narrow waists could only be shaped by high-speed collimated outflows or jets rather than pre-existing equatorial density enhancements.

With the arrival of the Hubble Space Telescope (HST) and state of the art CCD instruments in the mid 1990s, high-resolution images of PNe also revealed the widespread presence of microstructures such as bi- or multi-polar jets, knots, bubbles and filaments \citep{harrington94}.

Two major lines of investigation emerged into the origin of such microstructures. Single-star models attributed the collimation of outflows to strong magnetic fields which are fuelled by differential rotation in the stellar envelope. The claim was supported by three-dimensional magneto-hydrodynamic simulations which were successful in creating an impressive range of PN morphologies including multi-lobes and point symmetric PNe, which could not be accounted for by GISW theory \citep{dwarkadas98,garciasegura99,garciasegura00}. Recently, toroidal magnetic fields have been directly observed in the proto-PN W43A \citep*{vlemmings06} and the PN K3-35 \citep{miranda01,gomez09} through maser measurements, and in PNe NGC~7027, 6537, 6302 and proto-PN CRL~2688 through polarisation measurements \citep{greaves02,sabin07}, proving the existence of such fields. However, it has been argued that magneto-rotational launch, the mechanism for launching collimated jets with powerful magnetic fields, requires more energy than a single star can provide \citep{nordhaus07,frank06,soker04}. The star would initially be able to launch a jet but would not be able to sustain such magnetic fields for long periods of time, making the model inconsistent with PN evolution.

Collimated outflows are usually associated with quasars or X-ray binaries, where the accretion disc plays an essential role in feeding mass, energy and angular momentum into the jet. \citet{soker94} hence claimed that a stellar accretion disc must also be required to produce such outflows in PNe, thus inferring the presence of a binary companion. The outflow from a star in a close-binary system will also inevitably undergo a common-envelope (CE) stage during which the secondary is engulfed by the primary's expanding atmosphere into which it deposits energy and angular momentum through friction, giving the outflow a bipolar shape \citep*{lloyd97,porter98}.

In very recent years, binary systems have been considered to be essential to the formation of very elongated PNe. An accretion disc is a very efficient source of collimation \citep{soker01} and a companion star can spin-up the primary to help it sustain powerful magnetic fields via angular momentum convervation. Observational evidence of precessing jets in some PNe also implies the presence of binary central systems \citep*{lopez95}. Finally, recent stellar populations statistics carried out by \citet{moe06} showed that if both single stars and binaries evolve to PNe, the number of observed PNe is much lower than expected. The numbers are much closer if only binary stars are allowed to evolve to PNe, suggesting that binary systems are in fact essential to forming all PNe. However, currently only a very small fraction of PNe have confirmed close-binary central stars, likely to have produced accretion discs \citep{demarco09}. Estimations suggest that 10\%--15\% of all PNe harbour close binaries \citep{bond00}, and another $\sim10\%$ PNe with wider binaries (\citealt{ciardullo99}; see also the full discussion by \citealt{demarco09}).

The CE phase and jet collimation by accretion discs predict that a bipolar PN should be extended in the direction normal to the binary orbital plane, and also imply that any PN with a central close-binary should be non-spherical. Little observational work has been carried out on PNe harbouring close-binary central stars. We started in 2004 an observational campaign in order to obtain morpho-kinematical models of all PNe with confirmed close-binary central stars. We uncovered for the first time a direct link between PN shell and binary system after a common inclination angle was found in Abell 63 (A63) \citep{mitchell07}.

We present in this paper a kinematical study of the young PN Hubble 12 (Hb~12, R.A. 23:26:15, Dec. +58:10:55 [J2000.0]); an extremely well studied compact nebula ($\sim10\arcsec$ in diameter) which possesses a bipolar hourglass shell morphology \citep{sahai98}. It lies at a distance of 2.24 kpc \citep*{cahn92}. \citet{miranda89} performed a kinematic analysis of the nebula and found the inner shell to be expanding at $\sim16\kms\!$, and faint extended bipolar lobes. They also detected the presence of a faint, extended halo in \HA.

\citet*{hsia06} detected periodic photometric variability in the optical lightcurve of the central star and attributed it to an eclipsing binary system. However, doubt was cast on this interpretation of the observed photometric variability by \citet*{demarco08}, suggesting that there might not even be an eclipsing binary or that at least the binary parameters suggested were wrong.

Our aim is to construct a three-dimensional kinematical model of Hb~12, through the use of longslit spectroscopy, in order to determine its morphology and orientation. We will then be able to judge whether it is consistent with an eclipsing binary central star or not (assuming the binary is the main influence on the PN morphology). We also reveal for the first time direct observations of high-velocity knots in the system, both at a large distance and close to the central star, emitting in the optical; signs of strong outflow activity.

\section{Observations}\label{s2}

\subsection{Imaging}\label{s2_1}

\begin{figure*}
\begin{center}
\includegraphics[scale=0.9]{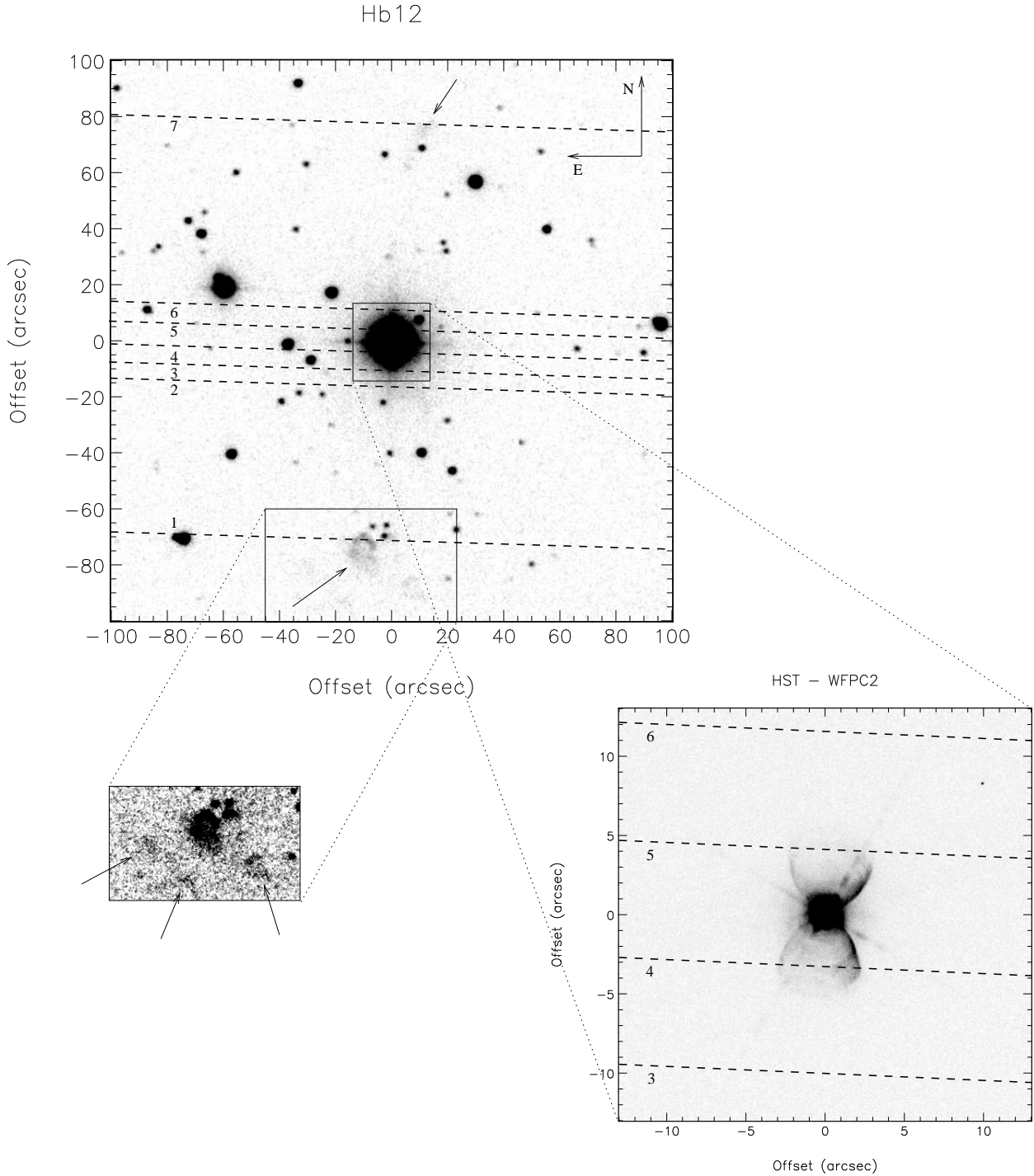}
\caption[Deep \NII image of Hb~12]{A deep \NII$\lambda6584$ image of Hb~12 taken with MES-SPM. North is up and east is left. The inset to the lower-right is an HST/WFPC2 \HA enlargement of the core revealing the hourglass morphology of Hb~12 (HST archive / \citealt{sahai98}). The seven E--W slit positions are marked with dashed lines. The arrows indicate the positions of the knots. The inset to the lower-left shows the region around the southern knot, revealing fainter knots, also indicated by arrows.}
\label{f1}
\end{center}
\end{figure*}

A deep 900 s narrowband \NII$\lambda6584$ image of Hb~12 was taken using the Manchester echelle spectrometer \citep{meaburn03} combined with the 2.1-m, f/8 San Pedro M\'{a}rtir telescope (MES-SPM) on $30^{\mathrm{th}}$ June 2000; this is shown in Fig.~\ref{f1}. The TEK--1 CCD was used with $1\times1$ binning ($\equiv0.31\arcsec$/pixel) and the seeing was 2\arcsec. A high resolution HST image taken by \citet{sahai98} is also displayed in lower-right inset in Fig.~\ref{f1}, showing the hourglass shape of the core of Hb~12.

The wide-field image unveils for the first time the presence of two knots on each side of the bright PN shell (indicated by the arrows) emitting in the low-ionisation \NII$\lambda6584$ line, aligned with the polar axis of the PN. The lower left-inset shows the region around the southern knot at a much higher contrast, in which three more much fainter knots (also indicated by arrows) begin to appear. Such knots are often found around PNe and are usually associated with high velocity outflows (see \citealt{lopez93,balick93,o'connor00,mitchell07}; for example). In analogy to other hourglass PNe, \citet{kwok07} predicted that outflow material should be present at a distance from the main nebula; our observations confirm this and are the first direct evidence for optical outflow material around Hb~12.

\citet{kwok07} observe in their infrared HST image of Hb~12 equatorial features associated with a second outer hourglass of a much bigger size and a wider opening angle, drawing a strong analogy to the spectacular nested hourglasses PN Hen~2-104 \citep{corradi01}. They also compare it to MyCn~18 in which a nested hourglass was also found \citep{sahai99}. The outer hourglass in Hb~12, along with the 'eye' created by the intersection of inclined bipolar lobes (which strongly resembles the central region of MyCn~18), are also visible in the $\mathrm{H}_{2}$ images of \citet{hora96}. However, as with \citet{kwok07}, we do not see any signs of this second hourglass in our optical image.

\subsection{Longslit spectroscopy}\label{s2_2}

Longslit spectroscopy of Hb~12 was performed at high spatial and spectral resolution using the MES-SPM on the nights of $10^{\mathrm{th}}$ November 1998 and $29^{\mathrm{th}}$ October 2000. Observations were taken through a narrow-band filter which isolates the \HA and \NII$\lambda6548$, $\lambda6584$ emission lines in the $87^{\mathrm{th}}$ echelle order. For the 1998 dataset, spectra were obtained using a TEK--1 CCD detector comprising $1024\times1024$ square pixels with sides measuring $24~\mu$m, using $2\times2$ binning ($\equiv0.62\arcsec$/pixel and 3.5 \kms/pixel). A TH2K CCD was used for the 2000 dataset with $2048\times2048$ square pixels with sides measuring $14~\mu$m, using $3\times3$ binning ($\equiv0.54\arcsec$/pixel and 3.1 \kms/pixel). Integration times were 1200 s and the seeing was 2\arcsec ~for all spectral observations.

A single slit with position angle $\mathrm{PA} = 86^{\circ}$ was placed across the nebula at seven different positions (see Fig.~\ref{f1}); positions 2--4 correspond to the data taken in 1998 whilst positions 1, 5, 6 and 7 belong to the 2000 dataset. For both epochs, the slit was 150 $\mu$m wide ($\equiv1.9\arcsec$ and 11 \kms) and 5\arcmin ~long. Slit positions 4 and 5 intersect the south and north lobes, respectively, of the bright hourglass core, and their \NII$\lambda6584$ position-velocity (p-v) arrays are shown in Fig.~\ref{f2}, row 1, columns \texttt{c} and \texttt{d}, respectively. The \HA p-v arrays from slit positions 2 to 6 are shown in Fig.~\ref{f3}. Slit positions 1 and 7 go through the knots to the south (only the brighter one) and north of Hb~12, respectively.

Data reduction was performed using the \textsc{starlink} software packages \textsc{kappa}, \textsc{figaro} and \textsc{ccdpack}. All images and spectra were bias-corrected and cleaned of cosmic rays. The \HA + \NII spectra were wavelength calibrated against a ThAr emission line lamp. All velocities have been corrected to the heliocentric frame of reference.

\section{Results}\label{s3}

\subsection{Measuring the inclination of the hourglass}\label{s3_1}

There is no doubt about the core morphology of Hb~12, the hourglass structure is clearly visible in the HST image inset in Fig.~\ref{f1}. However, the orientation of the shell is not clear. The common assumption for stellar outflows with binary central stars is that the orientation of the shell is close, if not identical, to that of the binary system (see \citealt{mitchell07} for example). \citet{hsia06} suggested that the central star system of Hb~12 was a partially eclipsing binary after detecting periodic variability in their photometry, which would limit the binary's inclination angle to $i_{\mathrm{binary}} > 80^{\circ}$. Yet, on the HST images, the orientation of the shell appears to be closer to $50^{\circ} - 70^{\circ}$, as shown by \citet{kwok07} who measured an inclination of $i_{\mathrm{shell}} = 52^{\circ}$ from fitting ellipses to co-axial rings observed in \NII HST images of Hb~12. Here we derive an inclination angle for the nebular shell based on the comparison of model spectra derived from an hourglass shaped model nebula with our observed \NII spectra from slits 4 and 5 which cross the two lobes of the hourglass.

\begin{figure*}
\begin{center}
\includegraphics[scale=0.965]{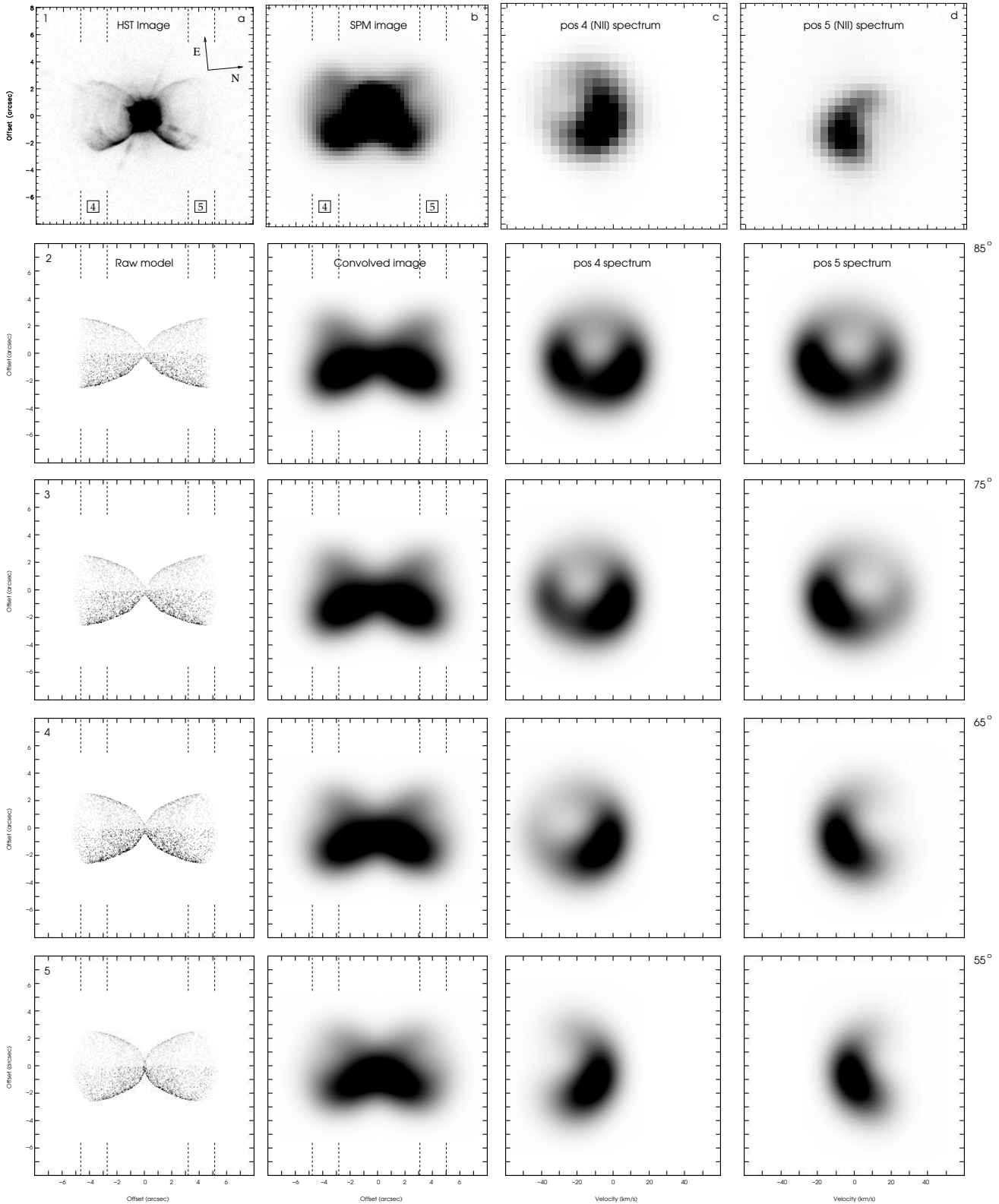}
\caption[Observed and synthetic \NII p-v arrays for slit positions 4 and 5]{\underline{Row 1}: optical data. (\texttt{a}) HST \HA image of the hourglass shell of Hb~12. (\texttt{b}) \NII$\lambda6584$ SPM image of Hb~12. (\texttt{c}) Position 4 \NII$\lambda6584$ p-v array. (\texttt{d}) Position 5 \NII$\lambda6584$ p-v array.\protect\\ \underline{Rows 2--5}: Column (\texttt{a}): Raw \textsc{shape} particle model. Column (\texttt{b}): raw model convolved with 2\arcsec ~seeing. Column (\texttt{c}): position 4 synthetic p-v array. Column (\texttt{d}): position 5 synthetic p-v array. Inclination angles are $85^{\circ}$, $75^{\circ}$, $65^{\circ}$ and $55^{\circ}$ for rows 2, 3, 4 and 5, respectively. The dashed lines in columns (\texttt{a}) and (\texttt{b}) mark the positions of slits 4 and 5.}
\label{f2}
\end{center}
\end{figure*}

We created a synthetic hourglass model of the shell using the \textsc{shape} code \citep{steffen06}. The hourglass morphology of the PN shell was based on the HST imagery of Hb~12 (inset in Fig.~\ref{f1} and the \NII images of \citealt{kwok07}) assuming axi-symmetry around the polar axis, and we gave it a Hubble-flow velocity law (proportional to distance from centre) combined with a systemic velocity of $-5~\kms$ \citep{hyung96} (see Fig.~\ref{f2}; column \texttt{a}). In the HST image, the PN shell is visibly brighter on the western side; this asymmetry in brightness was reproduced in the model, with a 300\% increase in brightness for the entire western half of the shell. The model was then convolved to the SPM image seeing of 2\arcsec ~(Fig.~\ref{f2}; column \texttt{b}), where the western brightening is now much more visible. We note here that extra bright emission is visible at the centre of the nebula in both the SPM and HST images, probably from the central star, which is not included in our model. However, this should not affect the following results since none of our slits positions go across this central region.

As an aside, it is interesting to note here that \citet{sahai98} found that the very core of Hb~12 (measuring $\sim 0.6\arcsec$) itself also shows signs of bipolarity with east-west inequalities in brightness, but unlike the PN shell, it is the eastern lobe of the core which is brighter than the western one; the origin of such anisotropy in brightness is not known. Note that the structure observed by \citet{sahai98} in the core is large compared to the size of a close-binary system: the centres of the two lobes, assuming a distance of 2.24 kpc, are separated by $\sim 350$ AU. In addition to this, the elongated nature of the lobes indicates that they are most probably not observing a resolved stellar binary.

Synthetic longslit p-v arrays were extracted from the model along slits matching positions 4 and 5 of the observations, convolved to the velocity resolution of the MES and thermal broadening of the \NII$\lambda6584$ line. Those were compared to the \NII$\lambda6584$ p-v arrays obtained from the longslit spectroscopy (Fig.~\ref{f2}; row 1, columns \texttt{c} and \texttt{d}). They were compared to the \NII$\lambda6584$ emission rather than \HA since Hb~12 is very bright in \HA and very little core structure is visible in the spectra at that wavelength, due to over-exposure. The \HA profiles were also contaminated by scattering effects from the very bright core of Hb~12. In addition, the \NII$\lambda6584$ line has an intrinsically narrower thermal width and tends to come from an outer `skin' whereas the \HA emission tends to be distributed through the volume, which also broadens the profiles.

Velocities of up to $-40~\kms$ and $+30~\kms$ are observed for positions 4 and 5, respectively, suggesting much higher expansion velocities than the $16~\kms$ quoted by \citet{miranda89} for the bright compact central region of the nebula. Our value for the Hubble-flow velocity scale factor was adjusted until the synthetic velocities from positions 4 and 5 matched the observed velocities; we retained a final value for the scale factor of 9.5 \kms/arcsec, and the resulting synthetic p-v arrays are shown in Fig.~\ref{f2} (columns \texttt{c} and \texttt{d}).

Our synthetic shell was tilted at various angles to the line of sight with the southern part of the shell towards the observer and the northern part away, and the results for $i_{\mathrm{shell}} = 85^{\circ}$, $75^{\circ}$, $65^{\circ}$ and $55^{\circ}$ are displayed in Fig.~\ref{f2} in rows 2, 3, 4 and 5, respectively. The model that best reproduces the observed p-v arrays is the $i_{\mathrm{shell}} = 65^{\circ}$ model (row 4). It becomes obvious from the synthetic p-v arrays in rows 2 and 3 that the PN shell is unlikely to have an inclination angle greater than $80^{\circ}$.

As mentioned above, \citet{kwok07} measured an inclination of $i_{\mathrm{shell}} = 52^{\circ}$ from fitting ellipses to co-axial rings observed in \NII HST images of Hb~12. \citet{hyung96} calculate $i_{\mathrm{shell}} = 40^{\circ}$ from optical spectroscopy and photoionisation models. In their kinematical study of the nested double hourglass Hen~2-104, \citet{corradi01} found that the inner and outer hourglasses had identical inclination angles. We fitted inclined circular rings to the equatorial features marked $b$ in the infrared image of \citet{kwok07} and found an inclination of $\sim60^{\circ}$ for the outer hourglass in Hb~12, which should be close, if not identical, to the inclination of its bright inner hourglass. Finally, \citet{hora96} also found $60^{\circ}$ from fitting an ellipse to the waist of the outer hourglass.

Current theories of PN evolution suggest that a binary stellar system will shape the ejected PN shell such that the nebula shares a common inclination angle with the binary system. Hence an eclipsing binary should produce a nebula with a symmetry axis at about $90^{\circ}$ (i.e. in the plane of the sky, see \citealt{mitchell07} for example).  All the measured values for the inclination of the hourglass shell of Hb 12 ($65^{\circ}, 60^{\circ}, 52^{\circ}$ and $40^{\circ}$), although not in strict agreement, together point to an inclination which is much lower than the $80^{\circ}$ limit which would be imposed if the central star of Hb~12 is indeed an eclipsing binary system. This inconsistency in inclination angle could possibly be explained if the central system was a reflection effect binary rather than a true eclipsing binary; \citet{hsia06} mention this as a possible interpretation of their lightcurve. One other way to maintain a common angle between the nebular shell and the progenitor system could be to invoke a triple system.

As mentioned earlier, there is also some doubt concerning the eclipsing binary in Hb~12 (see \citealt{demarco08}). In addition to the arguments presented by \citeauthor{demarco08}, the discrepancy between the binary and the nebular inclinations revealed by our study further suggests that the central stellar system is probably not eclipsing.

\subsection{Evidence for bipolar ejection}\label{s3_2}

\subsubsection{High-speed knots close to the central star}\label{s3_2_1}

\begin{figure}
\begin{center}
\includegraphics[scale=0.39]{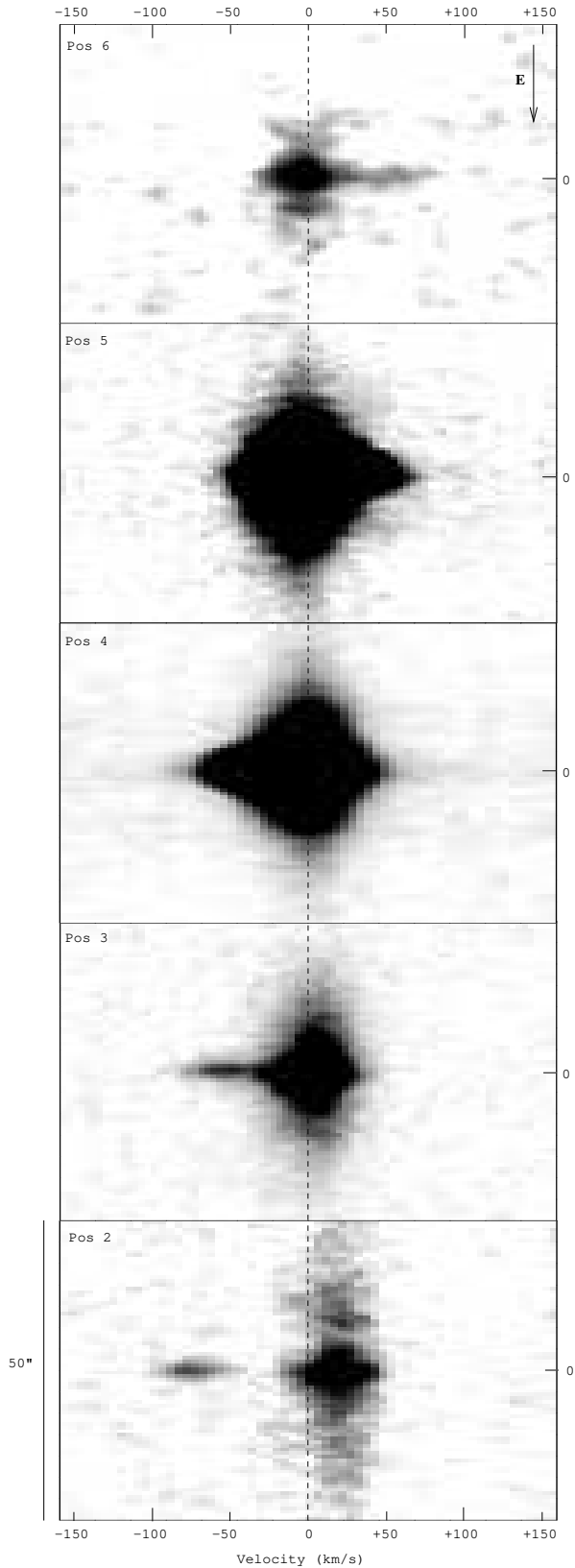}
\caption[Observed \HA p-v arrays for slit positions 2 to 6]{Position-velocity arrays for slit positions 2 to 6. Velocities are heliocentric. All spectra are \HA except for position 6 where the \NII$\lambda6584$ spectral line is shown, as we did not have an \HA spectrum at that slit position in our dataset. West is up and east is down and each frame measures 50\arcsec ~in height.}
\label{f3}
\end{center}
\end{figure}

In Fig.~\ref{f3}, we have displayed the \HA p-v arrays for slit positions 2 to 6, which are all close to the central star of Hb~12, using a common velocity axis. For each frame, the slit is vertical, west is at the top and east at the bottom. Each frame measures 50\arcsec ~in size in the vertical direction.

We can clearly identify a high-velocity ejection feature which is pointing in the direction away from the observer (positive velocity) for slit positions north of the central star of the PN (positions 5 and 6), and pointing towards the observer (negative velocity) for slits placed to its south (positions 2 to 4). This is the same north-south orientation as the PN shell. We also note that the velocity of the knots increase with distance from the central star. It is the first time that knots of this kind have been uncovered in Hb~12. \citet{welch99} detected \FEII shock excited regions along the polar direction close to the core that should naturally form in the cooling wake of the bow-shock produced by a jet or bullet, thus predicting the presence of ejecta in or around Hb~12; our observations confirm this for the first time.

\begin{figure}
\begin{center}
\includegraphics[scale=0.3]{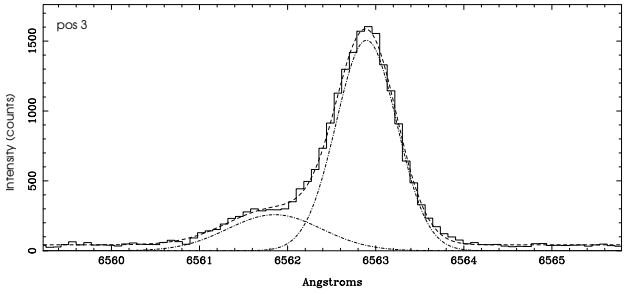}
\caption{Gaussian fits to the \HA emission line for position 3.}
\label{f4}
\end{center}
\end{figure}

We fitted Gaussian line profiles to the emission features in order to obtain measurements for the velocities of the knots; an example of this is shown in Fig.~\ref{f4}. The measured knot velocities (corrected for systemic motion) and angular distances from the central star are listed in Table~\ref{t1}. By assuming that the inclination of ejection direction is the same as that of the PN shell (i.e. $i_{\mathrm{ej}} = 65^{\circ}$) and that it runs down the polar axis of the hourglass, we computed the deprojected velocities at which the knots are moving away from the core, and their physical distances from the central star assuming a distance of 2.24 kpc.

\begin{figure}
\begin{center}
\includegraphics[scale=0.33]{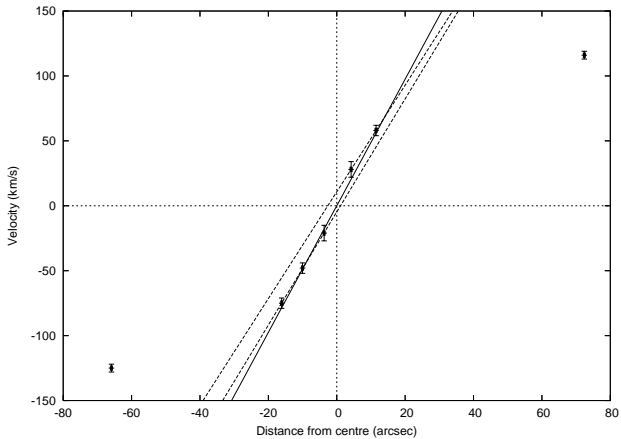}
\caption[Knot velocity as a function of angular distance from the central star]{Knot velocity as a function of observed angular distance from the central star. Velocities are heliocentric and have been corrected for the systemic motion of the PN. The straight line fit to the five data points close to the origin follows $V_{r} \propto \theta$. The dashed lines are separate fits to the data points above and below the origin. The two outer points correspond to the low-ionisation emission features at positions 1 and 7.}
\label{f5}
\end{center}
\end{figure}

We have plotted in Fig.~\ref{f5} the radial (systemic corrected) velocities of the knots as a function of angular distance from the centre of the PN. A straight line through the origin
\begin{equation}\label{eq1}
\displaystyle
\left(\frac{V_{r}}{\kms}\right) = (4.9\pm0.3) \times \left(\frac{\theta}{\mathrm{arcsec}}\right)
\end{equation}
adequately fits the five inner data points (solid line), consistent with Hubble-type expansion. Hence, the observed ballistic ejecta most probably originate from a single eruptive event similar to the ejecta observed in NGC~6302 \citep{meaburn05,meaburn08}.

We have also considered possible separate linear fits to the data points above and below the origin (dashed lines), similarly to the work of \citet{o'connor00} on MyCn~18, although many less data points are available to us. Such separate fits would suggest that the ejection of the material would have taken place at some distance on either side of the central star. It is interesting to note that this is the second hourglass PN for which this might be the case, even though no mechanism is currently known to produce such outflows. However, due to the uncertainties in the velocities and the low number of data points (for the points above the origin, we are `fitting' a straight line to only two data points!), we conclude that the simple single straight line fit is a much better description of this central outflow.

\begin{table*}
\begin{center}
\caption[Knot velocities in Hb~12]{Knot velocities, distances from centre and kinematical ages for the emission features observed in positions 1 to 7. The velocities have been corrected for systemic motion. Deprojected quantities and kinematical ages were calculated assuming $i_{\mathrm{ej}} = 65^{\circ}$ and $D = 2.24$ kpc.}
\begin{tabular}{cr@{$\pm$}lr@{$\pm$}lr@{$\pm$}lr@{$\pm$}lr@{$\pm$}l}
\hline
\hline
Slit     & \multicolumn{2}{c}{Radial  } & \multicolumn{2}{c}{Deprojected} & \multicolumn{2}{c}{Distance   } & \multicolumn{2}{c}{Deprojected       } & \multicolumn{2}{c}{Kinematical} \\
position & \multicolumn{2}{c}{velocity} & \multicolumn{2}{c}{velocity   } & \multicolumn{2}{c}{from centre} & \multicolumn{2}{c}{distance          } & \multicolumn{2}{c}{age        } \\
         & \multicolumn{2}{c}{$V_{r}$ } & \multicolumn{2}{c}{$V_{abs}$  } & \multicolumn{2}{c}{$\theta$   } & \multicolumn{2}{c}{$d$               } & \multicolumn{2}{c}{$KA$       } \\
         & \multicolumn{2}{c}{(\kms)  } & \multicolumn{2}{c}{(\kms)     } & \multicolumn{2}{c}{(arcsec)   } & \multicolumn{2}{c}{($\times10^{15}$m)} & \multicolumn{2}{c}{(years)    } \\
\hline
1        & $-125~$&$~3$                 & $-296~$&$~ 7$                   & $72.3~$&$~0.3$                  & $26.69~$&$~0.09$                       & $2860~$&$~ 60$                  \\
2        & $ -75~$&$~4$                 & $-178~$&$~10$                   & $16.1~$&$~0.3$                  & $ 5.95~$&$~0.09$                       & $1060~$&$~ 60$                  \\
3        & $ -48~$&$~4$                 & $-114~$&$~10$                   & $10.0~$&$~0.3$                  & $ 3.70~$&$~0.09$                       & $1030~$&$~ 90$                  \\
4        & $ -21~$&$~6$                 & $ -50~$&$~14$                   & $ 3.7~$&$~0.3$                  & $ 1.37~$&$~0.09$                       & $ 870~$&$~230$                  \\
5        & $ +28~$&$~6$                 & $ +66~$&$~14$                   & $ 4.2~$&$~0.3$                  & $ 1.56~$&$~0.09$                       & $ 750~$&$~100$                  \\
6        & $ +58~$&$~4$                 & $+137~$&$~10$                   & $11.5~$&$~0.3$                  & $ 4.25~$&$~0.09$                       & $ 980~$&$~ 70$                  \\
7        & $+116~$&$~5$                 & $+275~$&$~13$                   & $76.3~$&$~0.3$                  & $28.16~$&$~0.09$                       & $3250~$&$~ 90$                  \\
\hline
\end{tabular}
\label{t1}
\end{center}
\end{table*}

Finally, it is of course possible for the knots to have interacted with the circumstellar medium (CSM) which would most likely have caused them to decelerate, if the material is dense enough. The complex ejection of bursts of material at different velocities and/or different times into a CSM can only be properly treated using hydrodynamical models, and we cannot predict here the behaviour of the ejecta. However, we can show using a simple drag-force model (where the friction force acting on a moving clump is proportional to its velocity) that if the CSM is uniform and all the ejected clumps have the same mass and different ejection velocities, their position-velocity law would still go through the origin (Hubble-like, as in Fig.~\ref{f5}, solid line), thus rendering it indistinguishable from a case with no friction using different initial velocities. The gradient would only depend on the amount of friction (i.e. the density of the CSM) and the mass of the clumps. This scenario is somewhat idealistic and the probability for all these conditions to occur at once is rather low, which leads us to believe that the observed bullets would have only suffered a small amount of deceleration.

\subsubsection{High-speed external knots}\label{s3_2_2}

\begin{figure}
\begin{center}
\includegraphics[scale=0.45]{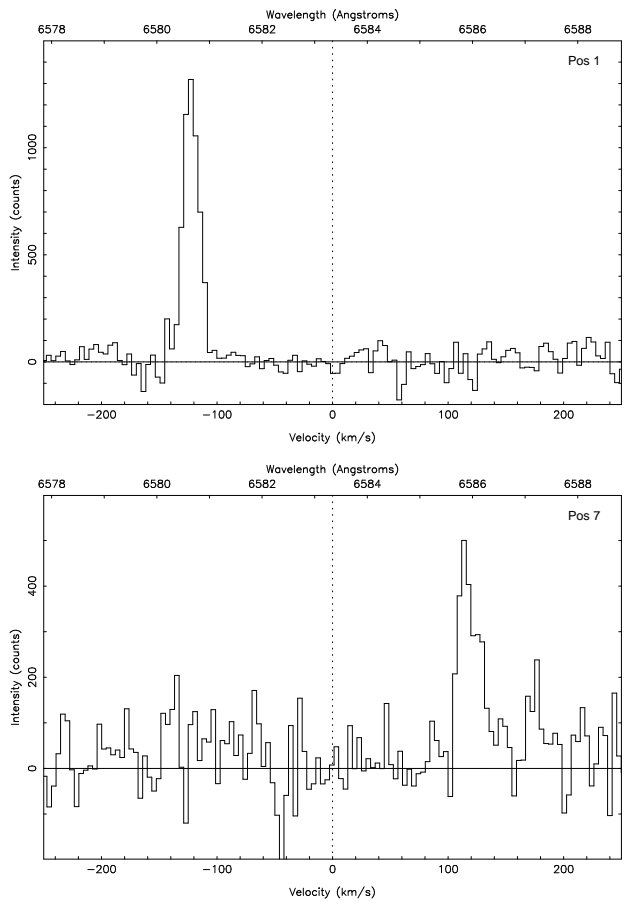}
\caption[\NII$\lambda6584$ spectra for slit positions 1 and 7]{\NII$\lambda6584$ spectra for slit positions 1 (top) and 7 (bottom) integrated over 10\arcsec ~in slit length. The velocities have been corrected for systemic motion.}
\label{f6}
\end{center}
\end{figure}

Slit positions 1 and 7 were placed on knotty emission regions visible in Fig.~\ref{f1}, and their \NII$\lambda6584$ spectra (integrated over $\sim10\arcsec$ ~of slit length) are shown in Fig.~\ref{f6}. The radial velocities were measured at $V_{r}(1) = -125\pm3~\kms$ and $V_{r}(7) = 116\pm3~\kms$ for positions 1 and 7, respectively. These end caps are high-speed low-ionisation bipolar ejecta along the hourglass axis, mostly visible in the \NII$\lambda6584$ emission line which implies collisional ionisation by shocks, usually associated with regions where fast ejecta meet the interstellar medium (ISM) (see \citealt{lopez93}). As for the inner knots, by assuming that the outer knots lie on the polar axis of the hourglass and that their ejection direction shares the same inclination angle, we calculated their deprojected velocities and distances; those are listed in Table~\ref{t1}.

The compact nature of the knots suggest that they were also formed by bullet-like ejection. However, when plotted alongside the inner bullet velocities (see Fig.~\ref{f5}), the outer points lie very far from the Hubble-type expansion law. If they all originated from the same eruptive event (as seems to be the case for the knotty outflows in NGC~6302; \citealt{meaburn05,meaburn08}), the outer ejecta must have suffered severe deceleration (more than 50\% of their initial velocity) from interaction with the ISM. This would also mean that the eruption ejected material on a very wide range of velocities $V\sim10-700~\kms$ (700 \kms ~would be the initial velocity of the outer knots if they followed the Hubble expansion law of the inner knots). The other possible explanations are that they belong to a different ejection event than the central ejecta and/or that they are at a different inclination angle.

\subsubsection{Kinematical ages}\label{s3_2_3}

The kinematical ages (KAs) for emission features 1 to 7 are listed in Table~\ref{t1}. The KAs for the end caps are approximately three times larger than for the inner emission features. It is thus likely that the inner and outer outflowing emitting gas belong to two separate eruptive events. The observation of separate eruptive events could be evidence that Hb~12 hosts a bipolar rotating episodic jet or BRET (\citealt*{lopez93}; \citealt{lopez95}), although it is not possible from our data to measure any rotation of the jet. One could also argue that the southern outer knot has a different KA than the northern one indicating that themselves could have come from separate eruptive events. However, at such large distances from the central star, deceleration effects from interactions with the ISM are more likely to be significant, leading to different values for KA for these knots even though they may have originated in the same event.

Finally, we calculate from our synthetic model of the hourglass nebular shell the KA of the nebula to be $1120\pm30$ years, which is of the same order as the KAs of the inner knots. This suggests that the end caps were formed prior to the PN shell, as was the case in A63 \citep{mitchell07}, most probably during the end of the AGB phase or the proto-PN phase. There is much evidence in the literature for bipolarity and ejection events in post-AGB stars (see \citealt{balick02}; for example).

This KA is much higher than the 300 years derived by \citet{miranda89}. The difference probably comes from the fact that their KA only applies to the bright central region of Hb~12 measuring 0.9\arcsec ~in diameter (which is not included in our model) compared to ours which applies to the entire bright hourglass. We suspect that these might in fact be two different components; the inner bright component may just be a bright thick stellar wind which is not kinematically part of the hourglass. Indeed, for a structure measuring 0.45\arcsec ~in radius, our hourglass Hubble law would give an expansion velocity of 4.3 \kms ~compared to the 16 \kms ~they used. Other differences might also come from the fact that the size of the component from which they derived their value was smaller than their seeing of 1.5\arcsec, even though they corrected for instrument and seeing effects (in fact, they point out that 0.9\arcsec ~is an upper limit and \citealt{sahai98}; measure it to be 0.6\arcsec, which would bring the KA down to $\sim200$ years for the same expansion velocity).

After studying many young PNe with point-symmetric morphologies, \citet{sahai98} have suggested an interesting evolutionary scenario. A collimated outflow (or several of them) from the central star system is initially active during the late AGB stage and carves structures in the surrounding AGB wind. This `sets the stage' for the development of an aspherical nebula where the hot fast wind (even if it is intrinsically spherical) runs into the imprinted AGB envelope. Such an evolution would be consistent with our calculated KAs.

\section{Discussion and conclusions}\label{s4}

We have created a morpho-kinematical model of Hb~12 based on imaging and high-resolution longslit spectroscopy of the nebula. We found that the shell inclination at which the synthetic spectra best matched the observations was $i_{\mathrm{shell}} = 65^{\circ}$, which is much lower than the inclination of a disputed eclipsing binary $i_{\mathrm{binary}} > 80^{\circ}$ derived from lightcurve studies. This result would go against current ideas of PN shaping from binary systems and the common binary-shell inclination as observed by \citet{mitchell07} in A63, and supports the proposition that if Hb~12 does host a binary central star, it is probably not eclipsing.

Further analysis of the \HA longslit spectra revealed the presence of a high-velocity bipolar outflows with Hubble-type velocities close to the PN core oriented in the same north-south direction as the PN shell. High velocities were also measured in the longslit spectra of the end caps observed in the wide-field image of Hb~12, showing they were also associated with jet-like activity. It is the first time that such outflows have directly been observed in Hb~12. Even though they were oriented in the same direction, large differences in velocities and KAs between the inner and outer knots led us to suspect that they originated from two separate eruptive events. We believe this may be evidence for a BRET in Hb~12.

Binary systems are believed to be essential to the formation of BRETs, where the accretion disc around the primary feeds matter and energy in the high velocity outflows, and precession or oscillations in the binary orbit cause the jets to precess likewise. This may suggest that Hb~12 does possess a non-eclipsing binary central star.

As noted by \citet{kwok07}, Hb~12 presents strong similarities to the hourglass nebulae Hen~2-104 and MyCn~18. The hourglass is first smooth close to the central star, and gets more irregular as we move away from the waist. The outer regions present a succession of rings, followed by some mottled structure where instabilities start to occur, with also some jet-like features emanating from the rim. We have also presented in this paper evidence for knotty outflows along the polar axis, as is the case for Hen~2-104 and MyCn~18. All three PNe present nested hourglass morphologies and the `eye' in the central regions of Hb~12 and MyCn~18 look very much alike. In the case of Hen~2-104, \citet{corradi01} found that the inner and outer nebulae had similar KAs, suggesting simultaneous ejection. With a lack of spectroscopic data on the outer hourglass, we cannot verify this in the case of Hb~12.

Hb~12 is the fourth object to have been discovered to have a nested hourglasses morphology (Hen~2-104, MyCn~18 and R~Aqr; see \citealt{solf85}), and in all three previous cases, the central system has been suggested or confirmed to be a symbiotic binary. \citet{kwok07} proposed that the central system in Hb~12 might also be a symbiotic binary (bearing in mind that there may be some doubt on their derived orbital parameters; \citealt{demarco08}). The hourglass PNe could be uncovering a link between PNe, symbiotic stars and novae (see also the discussion by \citealt{corradi03}; on nebulae around symbiotic binaries). Moreover, this link could be strengthened by studies reporting novae outbursts occuring inside old bipolar PNe with somewhat narrow waists. A faint PN-like shell was detected around the remnant of GK Per \citep*{bode04}, and more recently around V458 Vul \citep{wesson08}.

\section*{Acknowledgements}

NMHV acknowledges his University of Manchester research studentship during which most of the work presented here was performed, and his new ANR grant for the programme SiNeRGHy at the CEA/Saclay (ANR-06-CIS6-009-01). APR is supported by a STFC research studentship. JAL gratefully acknowledges financial support from DGAPA-UNAM grant IN116908. We would like to thank the referee for sensible comments which have improved this paper.

\label{lastpage}


\begin{thebibliography}{}

\bibitem[\protect\citeauthoryear{Balick \& Frank}{2002}]{balick02} Balick B., Frank A., 2002, ARA\&A, 40, 439

\bibitem[\protect\citeauthoryear{Balick, Preston \& Icke}{Balick et~al.}{1987}]{balick87} Balick B., Preston H.~L., Icke V., 1987, AJ, 94, 1641

\bibitem[\protect\citeauthoryear{Balick et~al.}{1993}]{balick93} Balick B., Rugers M., Terzian Y., Chengalur J.~N., 1993, ApJ, 411, 778

\bibitem[\protect\citeauthoryear{Bode, O'Brien \& Simpson}{Bode et~al.}{2004}]{bode04} Bode M.~F., O'Brien T.~J., Simpson M., 2004, ApJL, 600, L63

\bibitem[\protect\citeauthoryear{Bond}{2000}]{bond00} Bond H.~E., 2000, ASPCS, 199, 115

\bibitem[\protect\citeauthoryear{Cahn, Kaler \& Stanghellini}{Cahn et~al.}{1992}]{cahn92} Cahn J.~H., Kaler J.~B., Stanghellini L., 1992, A\&AS, 94, 399

\bibitem[\protect\citeauthoryear{Ciardullo et~al.}{1999}]{ciardullo99} Ciardullo R., Bond H.~E., Sipior M.~S., Fullton L.~K., Zhang C.-Y., Schaefer K.~G., 1999, AJ, 118, 488

\bibitem[\protect\citeauthoryear{Corradi}{2003}]{corradi03} Corradi R.~L.~M., 2003, ASP Conf. Ser., 303, 393

\bibitem[\protect\citeauthoryear{Corradi et~al.}{2001}]{corradi01} Corradi R.~L.~M., Livio M., Balick B., Munari U., Schwarz H.~E., 2001, ApJ, 553, 211

\bibitem[\protect\citeauthoryear{De Marco}{2006}]{demarco06} De Marco O., 2006, IAUS, 234, 111

\bibitem[\protect\citeauthoryear{De Marco}{2009}]{demarco09} De Marco O., 2009, ArXiv e-prints, 0902.1137

\bibitem[\protect\citeauthoryear{De Marco, Hillwig \& Smith}{De Marco et~al.}{2008}]{demarco08} De Marco O., Hillwig T.~C., Smith A.~J., 2008, AJ, 136, 323

\bibitem[\protect\citeauthoryear{Dwarkadas \& Balick}{1998}]{dwarkadas98} Dwarkadas V.~V., Balick B., 1998, ApJ, 497, 267

\bibitem[\protect\citeauthoryear{Frank}{2006}]{frank06} Frank A., 2006, IAUS, 234, 293

\bibitem[\protect\citeauthoryear{Garc\'{i}a-Segura et~al.}{1999}]{garciasegura99} Garc\'{i}a-Segura G., Langer N., R\'{o}\.{z}yczka M., Franco J., 1999, ApJ, 517, 767

\bibitem[\protect\citeauthoryear{Garc\'{i}a-Segura \& L\'{o}pez}{2000}]{garciasegura00} Garc\'{i}a-Segura G., L\'{o}pez J.~A., 2000, ApJ, 544, 336

\bibitem[\protect\citeauthoryear{G\'{o}mez et~al.}{2009}]{gomez09} G\'{o}mez Y., Tafoya D., Anglada G., Miranda L.~F., Torrelles J.~M., Patel N.~A., Hern\'{a}ndez R.~F., 2009, ApJ, 695, 930

\bibitem[\protect\citeauthoryear{Greaves}{2002}]{greaves02} Greaves J.~S., 2002, A\&AL, 392, L1

\bibitem[\protect\citeauthoryear{Harrington \& Borkowski}{2006}]{harrington94} Harrington J.~P., Borkowski, K.~J., 1994, BAAS, 26, 1469

\bibitem[\protect\citeauthoryear{Hora \& Latter}{1996}]{hora96} Hora J.~L., Latter W.~B., 1996, ApJ, 461, 288

\bibitem[\protect\citeauthoryear{Hsia, Ip \& Li}{Hsia et~al.}{2006}]{hsia06} Hsia C.~H., Ip W.~H., Li J.~Z., 2006, AJ, 131, 3040

\bibitem[\protect\citeauthoryear{Hyung \& Aller}{1996}]{hyung96} Hyung S., Aller L.~H., 1996, MNRAS, 278, 551

\bibitem[\protect\citeauthoryear{Kahn \& West}{1985}]{kahn85} Kahn F.~D., West K.~A., 1985, MNRAS, 212, 837

\bibitem[\protect\citeauthoryear{Kwok \& Hsia}{2007}]{kwok07} Kwok S., Hsia C.~H., 2007, ApJ, 660, 341

\bibitem[\protect\citeauthoryear{Livio et~al.}{1990}]{livio90} Livio M., Shankar A., Burkert A., Truran J.~W., 1990, ApJ, 356, 250

\bibitem[\protect\citeauthoryear{Livio \& Soker}{1988}]{livio88} Livio M., Soker N., 1988, ApJ, 329, 764

\bibitem[\protect\citeauthoryear{Lloyd, O'Brien \& Bode}{Lloyd et~al.}{1997}]{lloyd97} Lloyd H.~M., O'Brien T.~J., Bode M.~F., 1997, MNRAS, 284, 137

\bibitem[\protect\citeauthoryear{L\'{o}pez, Meaburn \& Palmer}{L\'{o}pez et~al.}{1993}]{lopez93} L\'{o}pez J.~A., Meaburn J., Palmer J.~W., 1993, ApJL, 415, L135

\bibitem[\protect\citeauthoryear{L\'{o}pez, V\'{a}zquez \& Rodr\'{i}guez}{L\'{o}pez et~al.}{1995}]{lopez95} L\'{o}pez J.~A., V\'{a}zquez R., Rodr\'{i}guez L.~F., 1995, ApJL, 455, L63

\bibitem[\protect\citeauthoryear{Meaburn et al.}{2003}]{meaburn03} Meaburn J., L\'{o}pez J.~A., Guti\'{e}rrez L., Quir\'{o}z F., Murillo J.~M., Vald\'{e}z J., Pedrayez M., 2003, Rev. Mex. de A. Ap., 39, 185

\bibitem[\protect\citeauthoryear{Meaburn et al.}{2005}]{meaburn05} Meaburn J., L\'{o}pez J.~A., Steffen W., Graham M.~F. Holloway A.~J., 2005, AJ, 130, 2303

\bibitem[\protect\citeauthoryear{Meaburn et al.}{2008}]{meaburn08} Meaburn J., Lloyd M., Vaytet N.~M.~H., L\'{o}pez J.~A., 2006, MNRAS, 385, 269

\bibitem[\protect\citeauthoryear{Miranda \& Solf}{1989}]{miranda89} Miranda L.~F., Solf J., 1989, A\&A, 214, 353

\bibitem[\protect\citeauthoryear{Miranda et~al.}{2001}]{miranda01} Miranda L.~F., G\'{o}mez Y., Anglada G., Torrelles J.~M., 2001, Nature, 414, 284

\bibitem[\protect\citeauthoryear{Mitchell et~al.}{2007}]{mitchell07} Mitchell D.~L., Pollacco D., O'Brien T.~J.,  Bryce M., L{\'o}pez J.~A., Meaburn J., Vaytet N.~M.~H., 2007, MNRAS, 374, 1404

\bibitem[\protect\citeauthoryear{Moe \& De Marco}{2006}]{moe06} Moe M., De Marco O., 2006, ApJ, 650, 916

\bibitem[\protect\citeauthoryear{Nordhaus, Blackman \& Frank}{Nordhaus et~al.}{2007}]{nordhaus07} Nordhaus J., Blackman E.~G., Frank A., 2007, MNRAS, 376, 599

\bibitem[\protect\citeauthoryear{O'Connor et al.}{2000}]{o'connor00} O'Connor J.~A., Redman M.~P., Holloway A.~J., Bryce M., L{\'o}pez J.~A., Meaburn J., 2000, ApJ, 531, 336

\bibitem[\protect\citeauthoryear{Porter, O'Brien \& Bode}{Porter et~al.}{1998}]{porter98} Porter J.~M., O'Brien T.~J., Bode M.~F., 1998, MNRAS, 296, 943

\bibitem[\protect\citeauthoryear{Sabin, Zijlstra \& Greaves}{Sabin et~al.}{2007}]{sabin07} Sabin L., Zijlstra A.~A., Greaves J.~S., 2007, MNRAS, 376, 378

\bibitem[\protect\citeauthoryear{Sahai \& Trauger}{1998}]{sahai98} Sahai R., Trauger J.~T., 1998, AJ, 116, 1357

\bibitem[\protect\citeauthoryear{Sahai et~al.}{1999}]{sahai99} Sahai R., et~al., 1999, AJ, 118, 468

\bibitem[\protect\citeauthoryear{Soker}{1997}]{soker97} Soker N., 1997, ApJ, 112, 487

\bibitem[\protect\citeauthoryear{Soker}{2001}]{soker01} Soker N., 2001, ApJ, 558, 157

\bibitem[\protect\citeauthoryear{Soker}{2004}]{soker04} Soker N., 2004, New Astronomy, 9, 399

\bibitem[\protect\citeauthoryear{Soker \& Livio}{1994}]{soker94} Soker N., Livio M., 1994, ApJ, 421, 219

\bibitem[\protect\citeauthoryear{Soker \& Rappaport}{2000}]{soker00} Soker N., Rappaport S., 2000, ApJ, 538, 241

\bibitem[\protect\citeauthoryear{Solf \& Ulrich}{1985}]{solf85} Solf J., Ulrich H., 1985, A\&A, 148, 274

\bibitem[\protect\citeauthoryear{Steffen \& L\'{o}pez}{2006}]{steffen06} Steffen W., L{\'o}pez J.~A., 2006, Rev. Mex. de A. Ap., 26, 30

\bibitem[\protect\citeauthoryear{Vlemmings, Diamond \& Imai}{Vlemmings et~al.}{2006}]{vlemmings06} Vlemmings W.~H.~T., Diamond P.~J., Imai H., 2006, Nature, 440, 58

\bibitem[\protect\citeauthoryear{Welch et~al.}{1999}]{welch99} Welch C.~A., Frank A., Pipher J.~L., Forrest W.~J., Woodward C.~E., 1999, ApJL, 522, L69

\bibitem[\protect\citeauthoryear{Wesson et~al.}{2008}]{wesson08} Wesson R., et~al., 2008, ApJL, 688, L21

\end{thebibliography}
\end{document}